\documentclass[aps,prc,twocolumn,floatfix,nofootinbib,showpacs]{revtex4}
\usepackage{amsmath,graphicx}

\newcommand{\mean}[1]{\left\langle #1 \right\rangle} 
\newcommand{\smean}[1]{\langle #1 \rangle} 

\newcommand{\ee}{{\rm e}}
\newcommand{\ii}{{\rm i}}

\begin{document}

\title{Azimuthally sensitive correlations in nucleus-nucleus collisions} 

\author{N. Borghini}
\email{borghini@spht.saclay.cea.fr}

\author{J.-Y. Ollitrault}
\email{Ollitrault@cea.fr}

\affiliation{Service de Physique Th{\'e}orique, CEA-Saclay,
  F-91191 Gif-sur-Yvette cedex, France}

\date{\today}

\begin{abstract}
We define a set of model-independent observables which generally 
characterize the azimuthal dependence of two-particle 
correlations in nucleus-nucleus collisions. We explain how they 
can be analyzed, and show to what extent such analyses are model
dependent. 
We discuss specific applications to the anisotropic flow of 
decaying particles, azimuthally sensitive HBT, 
and correlations between particles at large transverse momentum.
A quantitative prediction is made for jet quenching with respect
to the reaction plane. 
\end{abstract}


\maketitle

\section{Introduction}

In non-central nucleus-nucleus collisions, azimuthal angles 
of outgoing particles are generally correlated with the direction 
of the impact parameter. This phenomenon, called ``anisotropic flow,''
has been known for 20 years~\cite{Gustafsson:ka}, 
and has raised particular interest at the Brookhaven Relativistic 
Heavy Ion Collider (RHIC) where it is thought to provide unique 
evidence for quark-gluon plasma (QGP) formation~\cite{Gyulassy:2004zy}.

Most often, one studies the azimuthal dependence of {\it single-particle\/}
production~\cite{Reviews}. Here, we would like to discuss the
azimuthal dependence of {\it two-particle\/} correlations.
This is of interest in various situations:
\begin{itemize}
\item{{\it Anisotropic flow of short-lived particles\/}: 
  the flow of unstable particles (for instance $\Lambda$ baryons) is studied 
  through their decay products. One must first identify a correlation between 
  daughter particles, typically through an invariant mass plot; then study how 
  this correlation depends on the azimuthal angle of the decaying 
  particle~\cite{Justice:1997ma,Taranenko:1999,Chung:ny,Chung:2001je,%
    Barrette:2000cb,Adler:2002pb}.}
\item{{\it Azimuthally sensitive two-particle interferometry\/}: 
  Bose--Einstein correlations between identical particles are commonly used to 
  measure the size and shape of the emitting source~\cite{Wiedemann:1999qn}. 
  In non-central collisions, the source projection on the transverse plane is 
  no longer circular~\cite{Voloshin:1995mc}, and this can directly be seen in 
  Hanbury-Brown Twiss (HBT) studies of two-particle correlations, as already 
  observed at the Brookhaven Alternating Gradient Synchrotron~\cite{Lisa:2000xj} 
  and at  RHIC~\cite{Adams:2003ra}.}
\item{{\it Jet quenching with respect to the reaction plane\/}:  
  the energy loss of hard partons traversing a deconfined medium~\cite{%
    Bjorken:1982tu,Baier:1994bd} is a crucial signature of QGP formation at 
  RHIC~\cite{Wang:2004dn}. In particular, it results in a modification of the 
  pattern of azimuthal correlations between high-$p_T$ hadrons, compared to 
  $pp$ collisions: the back-to-back correlation is suppressed~\cite{%
    Adler:2002tq,Rak:2003ay}. In a non-central collision, the average length 
  of matter traversed by a parton depends on its azimuth~\cite{%
    Wang:2000fq,Lokhtin:jd}, which results in azimuthally dependent 
  two-particle correlations~\cite{Agakichiev:2003gg,Filimonov:2004qz}.}
\end{itemize}

In this paper, we give for the first time a unified presentation 
of these phenomena, which have so far been discussed separately. 
In all cases, analyzing azimuthally dependent correlations 
involves two distinct operations:
1) Measuring the distribution of a pair of particles with respect
to the reaction plane; 2) Isolating the ``true'' correlation 
from the uncorrelated part.
Both issues can be discussed independently, on a fairly general
footing. 

The first operation is discussed in Sec.~\ref{s:definitions}, 
where the observables associated with two-particle anisotropic 
flow are defined. 
These observables are model-independent and can in principle
be measured accurately.
In particular, it will clearly appear that any method used to measure 
the single-particle anisotropic flow can also be used to analyze 
azimuthally sensitive correlations, modulo minor modifications. 
In that view, we recall in Appendix~\ref{s:appendix} the main features 
of existing methods for analyzing one-particle flow, and we introduce the 
changes necessary to measure pair flow. 
While existing methods all require to estimate the reaction plane on an 
event-by-event basis~\cite{Danielewicz:hn,Poskanzer:1998yz} (see for instance
Refs.~\cite{Sorensen:2003kp} for $\Lambda$ flow, \cite{Heinz:2002au} for 
azimuthally sensitive HBT, and \cite{Bielcikova:2003ku} for correlations 
between high-momentum particles), this step is by no means necessary with 
the procedure we suggest. 
This opens the possibility to apply the improved methods of flow analysis 
recently devised in Refs.~\cite{Borghini:2001vi,Borghini:2001zr,%
  Borghini:2002vp,Bhalerao:2003xf,Borghini:2004ke} and to resolve an 
inconsistency of present analyses: on the one hand, one studies a correlation 
(between decay products, due to quantum statistics, from jet fragmentation) 
which is essentially a ``nonflow'' correlation; on the other hand, one uses 
the event-plane method which relies on the assumption that all correlations 
between particles are due to flow~\cite{Danielewicz:hn}. 

The second operation is discussed in Sec.~\ref{s:correlation}. 
Unlike the first one, it will be shown to be always model-dependent. 
Several specific applications are discussed in Sec.~\ref{s:applications}, 
together with predictions regarding the pair-flow coefficients. 
Our results are summarized in Sec.~\ref{s:conclusions}.

\section{Observables for two-particle anisotropic flow}
\label{s:definitions}

We first recall definitions for single-particle distributions. 
For particles of a given type in a given 
rapidity ($y$) and transverse momentum ($p_T$) window, the 
probability distribution of the azimuthal angle $\phi$ 
(measured with respect to a fixed direction in the laboratory)
reads 
\begin{equation}
\label{1p-dist}
p(\phi-\Phi_R)=\frac{1}{2\pi}
\sum_{n=-\infty}^{+\infty} v_n\,\ee^{\ii n(\phi-\Phi_R)}, 
\end{equation}
where $\Phi_R$ is the (unknown) 
azimuth of the reaction plane (impact parameter) in the laboratory frame. 

The Fourier coefficients~\cite{Voloshin:1994mz} in this expansion are 
given by $v_n=\smean{\ee^{-\ii n(\phi-\Phi_R)}}$, where angular brackets 
denote an average over particles and events. 
Given the normalization choice in Eq.~(\ref{1p-dist}), $v_0=1$. 
Since $p(\phi-\Phi_R)$ is real, $v_{-n}=(v_n)^*$, where $^*$ 
denotes the complex conjugate. 
If, in addition, the system is symmetric with respect to the reaction plane 
[$-(\phi-\Phi_R)$ is equivalent to $\phi-\Phi_R$], as in a collision between 
spherical (although not necessarily identical) nuclei when parity is 
conserved, Eq.~(\ref{1p-dist}) reads
\begin{equation}
\label{1p-dist2}
p(\phi-\Phi_R)=\frac{1}{2\pi}\!
\left[1+2\sum_{n=1}^{+\infty} v_n\cos n(\phi-\Phi_R) \right]
\end{equation}
with $v_n=\mean{\cos n(\phi-\Phi_R)}$, i.e., $v_n$ is real. 

The Fourier coefficients $v_n$ have by now become familiar in the study 
of anisotropic flow in (ultra)relativistic heavy-ion collisions. 
Nevertheless, it is instructive to recall why they are the proper tools 
to parameterize azimuthal anisotropies.
The key feature is that even though the reaction plane $\Phi_R$ is unknown 
on an event-by-event basis, the first Fourier coefficients $v_n$ can be 
accurately reconstructed from a statistical analysis of azimuthal 
correlations between outgoing particles (see Appendix~\ref{s:methods1} for 
a review of the methods for analyzing single-particle flow). 
However, the higher the value of $n$, the larger the 
uncertainty on $v_n$~\cite{Ollitrault:1997di}. 
Therefore the probability $p(\phi-\Phi_R)$ at a specific azimuth
cannot be measured in practice. 
Furthermore, since $v_n$ is defined as an average, it is also 
easier to compute in theoretical studies---in particular, in Monte 
Carlo models--- than the probability distribution itself. 

\begin{figure}[tb!]
  \centerline{\includegraphics*[width=0.5\linewidth]{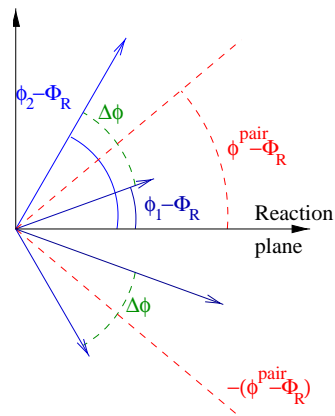}}
  \caption{Illustration of the various azimuthal angles $\phi_1$, $\phi_2$, 
    $\phi_{\rm pair}$, $\Delta\phi$, with $x=\frac{1}{2}$. 
    \label{fig:angles}}
\end{figure}

The above definitions can readily be generalized to the distribution 
of particle {\it pairs\/} with respect to the reaction plane. 
A pair of particles of given species is characterized by 
6 kinematic variables ${p_T}_1,y_1,\phi_1,{p_T}_2,y_2,\phi_2$. 
It is convenient to combine $\phi_1$ and $\phi_2$ into the relative angle 
$\Delta\phi\equiv\phi_2-\phi_1$ (or any similar observable that does not 
depend on the overall orientation of the pair in the transverse plane, 
as e.g. the invariant mass) and a ``pair angle'' 
\begin{equation}
\label{defPhi}
\phi_{\rm pair}\equiv x\phi_1+(1-x)\phi_2,
\end{equation}
where $0\le x\le 1$. 
One can restrict $\phi_{\rm pair}$ and $\Delta\phi$ to the ranges 
$-\pi\leq\phi_{\rm pair}<\pi$ and $-\pi\leq\Delta\phi<\pi$. 
If $x=\frac{1}{2}$, $\phi_{\rm pair}$ is the mean angle. 
The choice of $x$ depends on the problem under study: 
most often, one chooses for $\phi_{\rm pair}$ the azimuthal 
angle of the total transverse momentum ${{\bf p}_T}_1+{{\bf p}_T}_2$
(see Sec.~\ref{s:resonances} and Sec.~\ref{s:HBT}); 
in studies of azimuthal correlations between high-momentum 
particles, $x=1$ is a more common choice (see Sec.~\ref{s:jets}). 

Consider now a sample of pairs of particles in some range of 
${p_T}_1$, ${p_T}_2$, $y_1$, $y_2$, $\Delta\phi$. 
To study the probability distribution of the pair angle $\phi_{\rm pair}$ 
within this sample, we write its probability distribution in a way 
analogous to Eq.~(\ref{1p-dist}): 
\begin{equation}
\label{2p-dist}
p(\phi_{\rm pair}-\Phi_R)=\frac{1}{2\pi}
\sum_{n=-\infty}^{+\infty}v_n^{\rm pair}\,\ee^{\ii n(\phi_{\rm pair}-\Phi_R)}. 
\end{equation}
As the usual $v_n$'s, the ``pair-flow'' coefficients $v_n^{\rm pair}$ are 
defined by $v_n^{\rm pair}=\smean{\ee^{-\ii n(\phi_{\rm pair}-\Phi_R)}}$, 
with the normalization $v_0^{\rm pair}=1$. 
Since the probability distribution is real-valued, 
the coefficients also satisfy the property 
\begin{equation}
\label{property1}
v_{-n}^{\rm pair}=\left(v_{n}^{\rm pair}\right)^*.
\end{equation}
But unlike the single-particle flow $v_n$, the pair-flow coefficient
$v_n^{\rm pair}$ is in general not a real number. 
The underlying reason, exemplified in Fig.~\ref{fig:angles}, is that the 
transformation $\phi_{\rm pair}-\Phi_R \to -(\phi_{\rm pair}-\Phi_R)$ 
for a constant $\Delta\phi$ 
is {\em not\/} a symmetry of the system.\footnote{%
  The actual symmetry is under the simultaneous transformation 
  $\phi_{\rm pair}-\Phi_R \to -(\phi_{\rm pair}-\Phi_R)$, 
  $\Delta\phi\to -\Delta\phi$. Its consequences for the coefficients 
  $v_n^{\rm pair}$ are discussed in Appendix~\ref{s:appendixB}.}
As a consequence, sine terms are also present in the real form of the 
Fourier expansion, and Eq.~(\ref{1p-dist2}) is replaced by 
\begin{eqnarray}
\label{2p-dist2}
p(\phi_{\rm pair}-\Phi_R)&\!\!\equiv\!\!&\frac{1}{2\pi}
\Bigg(\! 1+ 2\!\sum_{n=1}^{+\infty}\!\left[ 
    v_{c,n}^{\rm pair}\cos n(\phi_{\rm pair}\!-\!\Phi_R) \right. \cr
 & & \qquad\qquad\left.
   +\,v_{s,n}^{\rm pair}\sin n(\phi_{\rm pair}\!-\!\Phi_R) \right]\!\!\Bigg)\!,
 \quad
\end{eqnarray}
where the {\em real\/} coefficients 
$v_{c,n}^{\rm pair} = \smean{\cos n(\phi_{\rm pair}-\Phi_R)}$ and 
$v_{s,n}^{\rm pair} = \smean{\sin n(\phi_{\rm pair}-\Phi_R)}$ 
are related to the complex $v_n^{\rm pair}$ by the relation 
$v_n^{\rm pair}=v_{c,n}^{\rm pair}-\ii v_{s,n}^{\rm pair}$.

The existence of such sine terms was already noted in 
Ref.~\cite{Heinz:2002au} in the context of azimuthally sensitive 
HBT studies. Note that it does not imply parity
violation, as it would in the case of single-particle
flow~\cite{Voloshin:2000xf}. 
The physical meaning of these additional terms will be illustrated
in Sec.~\ref{s:applications} in various physical situations. 
In particular, we shall show that they may yield insight on the 
mechanism responsible for the deficit in high-$p_T$ particles. 

It may be interesting to note that the pair-flow coefficients 
$v_n^{\rm pair}(\Delta\phi)$, when viewed as functions of the relative 
angle $\Delta\phi$, have peculiar properties which are derived in 
Appendix~\ref{s:appendixB}. 
Checking that measured values of the coefficients possess these properties 
then provides a way to evaluate the errors affecting the measurement. 

In experimental analyses, any method that can be used to measure the 
single-particle flow $v_n$ can be applied to extract 
the cosine terms $v_{c,n}^{\rm pair}$, without any modification:
one simply considers the pair as a single particle with azimuthal
angle $\phi_{\rm pair}$. The generalizations required in order to extract 
the sine terms $v_{s,n}^{\rm pair}$ are quite straightforward. 
They are summarized in Appendix~\ref{s:methods2} for various methods 
of flow analysis. 
In particular, some of these methods can safely correct for nonflow 
effects, others for acceptance anisotropies. 
This extends the possibility of studying azimuthally-dependent 
correlations to detectors with partial azimuthal coverage.

To conclude this Section, let us emphasize that the new characterization of 
azimuthally-sensitive two-particle correlations which we propose, with 
pair-flow Fourier coefficients, represents in our view an improvement over 
previous parameterizations, in the same way as $v_n$ is an improvement 
over older observables for anisotropic flow. 
The reason is simply that $v_{c,n}^{\rm pair}$ and 
$v_{s,n}^{\rm pair}$ are model-independent and detector-independent 
observables.

\section{Isolating the correlated part}
\label{s:correlation}

Subtracting the ``trivial'' uncorrelated part in order to 
isolate the ``true'' correlation is far from trivial. 
In this Section, we discuss this issue in as simple and general
a way as possible. 
For the sake of simplicity, we start with the case when there is 
no anisotropic flow. 
In Sec.~\ref{s:correlationA}, we 
explain why the subtraction always involves some 
degree of arbitrariness, most often in the form of an 
arbitrary constant. 
Although this is to some degree well-known, at least to 
those who actually perform correlation analyses, we think it is
worth recalling, since the literature on the subject is rather 
confusing. 
In Sec.~\ref{s:correlationB}, we 
recall the various ways of normalizing the correlation, depending 
on the observable under study. 

In practice, however, anisotropic flow is most often present, 
and makes the background subtraction more difficult in 
heavy ion collisions than in elementary collisions. 
Note that this applies to all correlation analyses, 
not only to the azimuthally-dependent ones: 
the correlation of single particles with the reaction plane 
induces a correlation between them, which must be always 
subtracted, at least in principle, in order to isolate 
other effects (see for instance \cite{Adler:2002ct}).  
In Sec.~\ref{s:correlationC}, we explain how this can be done, 
and show that this subtraction implies further approximations. 


\subsection{A model-dependent issue}
\label{s:correlationA}

In a given event, let $N_1$ and $N_2$ denote the numbers 
of particles in two phase-space bins, $(p_{T1},y_1,\phi_1)$ and 
$(p_{T2},y_2,\phi_2)$. 
To simplify the discussion, we assume that the two bins are separated. 
If they overlap, one need only replace $N_1N_2$ by the 
number of pairs in what follows. 
The simplest definition of the correlation 
between the two bins is 
\begin{equation}
\label{defcorr1}
{\cal C}=\mean{N_1N_2}-\mean{N_1}\mean{N_2}, 
\end{equation}
where angular brackets mean an average over many 
events.\footnote{In other terms, $\mean{N_1}$, $\mean{N_2}$ 
  and $\mean{N_1N_2}$ are the one- and two-particle inclusive 
  cross-sections, divided by the total inelastic nucleus-nucleus 
  cross-section.}

Such a definition is not satisfactory in practice 
because the sample of events used in the analysis 
always contains events with different centralities: 
in particular, the total multiplicity may have sizeable
fluctuations within the sample of events considered, 
and these fluctuations alone induce a correlation between 
any two phase-space bins. This correlation is of a rather 
trivial nature, but it may well overwhelm the interesting 
ones~\cite{Carruthers:1990dn}. 
A simple way out of this problem would be to normalize the 
two terms in Eq.~(\ref{defcorr1}) by the total number of pairs 
of correlated and uncorrelated particles, respectively, that is, to 
define instead the correlation as 
\begin{equation}
\label{defcorr2}
{\cal C}=\mean{N_1N_2}-\frac{\mean{N(N-1)}}{\mean{N}^2}
\mean{N_1}\mean{N_2},
\end{equation}
where $N$ is the number of particles in a large enough phase-space 
bin (typically, the total number of charged pions seen in the detector 
for interferometry analyses). 
This definition is also unsatisfactory for two reasons: 
first, it obviously introduces some degree of arbitrariness
in the definition of the correlation, depending on the 
choice of the phase space for $N$; second, part of the 
fluctuations in $N$ may be meaningful 
for the correlation analysis, as in the case of 
Bose--Einstein correlations~\cite{Miskowiec:1997ay}, so that 
there is no point in subtracting them. 

In actual analyses, the correlation is rather defined as 
\begin{equation}
\label{defcorr3}
{\cal C}=\mean{N_1N_2}-c\mean{N_1}\mean{N_2},
\end{equation}
where $c$ is some free coefficient. 
This coefficient is kept constant throughout the correlation analysis 
(which typically involves varying the invariant mass, the 
relative momentum, or the relative azimuth between the two particles). 
It is then fitted in such a way that the correlation ${\cal C}$ vanishes 
when it is expected to: at large relative momentum in HBT 
analyses~\cite{Miskowiec:1997ay}, 
in some range of $\Delta\phi$ in correlations between 
high-$p_T$ particles~\cite{Adler:2002tq,Rak:2003ay}. 
In addition,  as mentioned above, the background subtraction is often 
complicated by the existence of anisotropic flow, as we shall
discuss in Sec.~\ref{s:correlationC}. 

\subsection{Normalizations}
\label{s:correlationB}

There are essentially three ways of normalizing yields of particle 
pairs in azimuthally-independent analyses, depending on the observable under 
study. 
\begin{enumerate}
\item{One simply computes the average number of pairs per event, 
$\smean{N_1N_2}$. 
For instance, in order to measure $\Lambda$ production, 
one plots the number of $(p,\pi^-)$ pairs per event as a function 
of the invariant mass $M$ of the pair. The number of pairs in the
peak around the $\Lambda$ mass 
gives the yield of $\Lambda$ baryons, modulo acceptance
corrections.}
\item{One divides the number of pairs $\smean{N_1N_2}$ 
by the number of uncorrelated pairs, $c\smean{N_1}\smean{N_2}$. 
This is the standard observable for Bose--Einstein correlations, 
where the ratio varies ideally between 2 and 1 as the relative 
momentum of the pair increases.}
\item{The third, intermediate choice is to divide the number of pairs 
$\smean{N_1N_2}$ by the number of ``trigger particles'' 
$\smean{N_1}$~\cite{Adler:2002tq}. After subtraction of the 
uncorrelated part $c\smean{N_2}$, one thus obtains the mean 
number of particles $N_2$ correlated  with a trigger particle, 
which is independent of the system size (i.e., the same for 
a nucleus-nucleus and for a proton-proton collision) if there 
is no final-state interaction. }
\end{enumerate}

\subsection{Subtracting the correlation due to flow}
\label{s:correlationC}

When the particles in the pair are individually correlated with the 
reaction plane $\Phi_R$, this induces a trivial correlation between 
them, which must also be subtracted.

This subtraction is easy in principle: one simply repeats the operations 
of Secs.~\ref{s:correlationA} and \ref{s:correlationB} for a fixed 
orientation of the reaction plane $\Phi_R$. 
Then, the following substitutions hold 
\begin{eqnarray}
\label{subst}
\mean{N_1N_2}&\to&\mean{N_1N_2}(2\pi)\,p(\phi_{\rm pair}-\Phi_R)\cr
\mean{N_1}&\to&\mean{N_1}(2\pi)\,p_1(\phi_1-\Phi_R)\cr
\mean{N_2}&\to&\mean{N_2}(2\pi)\,p_2(\phi_2-\Phi_R). 
\end{eqnarray}
In these equations, $\mean{N_1N_2}$, $\mean{N_1}$ and $\mean{N_2}$ 
denote quantities averaged over $\Phi_R$; 
$p(\phi_{\rm pair}-\Phi_R)$ is the distribution of the pair angle, defined 
in Eq.~(\ref{2p-dist2}), and $p_1(\phi_1-\Phi_R)$, $p_2(\phi_2-\Phi_R)$ 
denote the single-particle azimuthal distributions of each particle, 
defined as in Eq.~(\ref{1p-dist2}). 

Once the azimuthal distributions of pairs and single particles 
with respect to the {\em true\/} reaction plane have been properly 
reconstructed, extracting the correlation, and its azimuthal dependence, 
is straightforward. 

Strictly speaking, however, the one- and two-particle probabilities 
$p(\phi-\Phi_R)$ and $p(\phi_{\rm pair}-\Phi_R)$ at a specific azimuth 
relative to the reaction plane cannot be reconstructed. 
As already mentioned in Sec.~\ref{s:definitions}, only the first few 
Fourier coefficients $v_n$ or $v_n^{\rm pair}$ can be reconstructed, due 
to larger absolute uncertainties on higher-order coefficients. 
On the other hand, the Fourier coefficients of a smooth function 
of $\phi-\Phi_R$ are expected to decrease quickly as the order increases 
(this expectation is supported by recent experimental single-particle flow
data \cite{Adams:2003zg}), so that one can reasonably truncate the series, 
keeping only the measured coefficients. 
This truncation is always required in order to estimate the correlation 
{}from anisotropic flow~\cite{Adler:2002tq,Bielcikova:2003ku}. 
At RHIC, for instance, the error on the azimuthal distribution at 
midrapidity is likely to be dominated by the error on the fourth 
harmonic $v_4$, and one can take $\frac{1}{\pi}\delta v_4$ as the 
error on $p(\phi-\Phi_R)$.

\section{Applications}
\label{s:applications}

We shall now discuss specific applications, with emphasis on the 
details of the experimental procedure. 

\subsection{Anisotropic flow of short-lived particles}
\label{s:resonances}

Let us begin with the measurement of the anisotropic flow of particles that 
are seen through their decay products, such as $\Lambda\to p\pi^-$~\cite{%
  Justice:1997ma,Chung:2001je,Barrette:2000cb,Adler:2002pb}, 
$\pi^0\to\gamma\gamma$, $\eta\to\gamma\gamma$~\cite{Taranenko:1999},
$K_S^0\to\pi^+\pi^-$~\cite{Chung:ny,Adler:2002pb}. 
We shall illustrate the recipe by discussing the flow of 
$\Lambda$ baryons. 

For each event, one sorts $(\pi^-,p)$ pairs into bins of invariant 
masses $M$.
The first step is then to analyze the total $\Lambda$ yield:
following the standard procedure, one counts the number of pairs 
in each invariant-mass bin, irrespective of the pair azimuth; 
let $N_{\rm pairs}(M)$ denote this number. 
One then separates this distribution into an uncorrelated part 
[the background $N_b(M)$] and a correlated part [the peak 
$N_\Lambda(M)$, centered around the expected $\Lambda$ mass]:
\begin{equation}
\label{background}
N_{\rm pairs}(M)=N_b(M)+N_\Lambda(M). 
\end{equation}
The integral of the correlated part $N_\Lambda(M)$ over 
$M$ is the $\Lambda$ yield. 

In most cases, the peak is well above the background: to perform the 
above decomposition, one need not go through the whole procedure 
of the previous section: instead, one simply assumes that the 
background $N_b(M)$ is a smooth function of $M$~\cite{Sorensen:2003kp}.
Please note that the anisotropic flow of $\pi^-$ and $p$ correlates 
their azimuthal angles, and therefore distorts the background. 
However, the distorted background remains smooth, so that this effect
need not be taken into account.

Next, one defines the azimuthal angle of the pair, $\phi_{\rm pair}$, 
as the azimuthal angle of the total transverse momentum 
${{\bf p}_T}_1+{{\bf p}_T}_2$, and one analyzes the pair flow coefficients 
$v_{c,n}^{\rm pair}(M)$ and $v_{s,n}^{\rm pair}(M)$ in each bin.
One then performs a decomposition similar to Eq.~(\ref{background})
for the azimuthally dependent part of the pair yield:
\begin{eqnarray}
\label{backgroundflow}
N_{\rm pairs}(M)\,v_{c,n}(M)&\!\!=\!\!&N_b(M)\,v_{c,n}^{(b)}(M)+
N_\Lambda(M)\,v_{c,n}^\Lambda\cr
N_{\rm pairs}(M)\,v_{s,n}(M)&\!\!=\!\!&N_b(M)\,v_{s,n}^{(b)}(M)+
N_\Lambda(M)\,v_{s,n}^\Lambda.\nonumber\\[-1mm]
 & &
\end{eqnarray}
This decomposition is performed assuming that the background components 
$N_b(M)\,v_{c,n}^{(b)}(M)$, $N_b(M)\,v_{s,n}^{(b)}(M)$ 
are smooth functions of $M$.
In this particular case, symmetry with respect to the reaction plane for 
$\Lambda$ particles implies $v_{s,n}^\Lambda=0$, except for
experimental biases and fluctuations.  
This identity can be used in order to check the accuracy of the 
experimental procedure, as in the case of single-particle 
flow~\cite{Poskanzer:1998yz}.
If the background consists of uncorrelated particles, one also has 
$v_{s,n}^{(b)}(M)=0$. 

In most analyses so far, the decomposition between the background 
and the peak is performed independently for several bins (typically, 20) 
in $\phi_{\rm pair}-\Psi_R$ 
\cite{Adler:2002pb,Barrette:2000cb,Sorensen:2003kp}, where 
$\Psi_R$ is an estimate of the reaction plane. 
With the above procedure, the decomposition is only performed 
twice, in Eqs.~(\ref{background}) and (\ref{backgroundflow}).

When the peak/background ratio is low, finally, mixed events 
can be used to define the background~\cite{Taranenko:1999}. 
However, the above-mentioned distortion of the background 
due to anisotropic flow must then be taken into 
account, as we shall see in more detail in Sec.~\ref{s:HBT}.

\subsection{Quantum correlations}
\label{s:HBT}

Azimuthally dependent Bose--Einstein (or, more generally, short-range)
correlations are analyzed in two steps. 
The first step is to perform a Fourier expansion of the pair 
yield with respect to the reaction plane, 
for each relative momentum ${\bf q}$~\cite{Heinz:2002au}. 
As explained in Sec.~\ref{s:definitions}, any method of flow analysis 
can be used to extract the Fourier coefficients 
$v_{c,n}({\bf q})$ and $v_{s,n}({\bf q})$. 
Even with the event-plane method, no binning in 
$\phi_{\rm pair}-\Psi_R$ is required, in 
contrast to present analyses~\cite{Lisa:2000xj,Adams:2003ra}. 
Once the coefficients are known, one easily builds the distribution of pairs 
relative to the reaction plane (up to truncation issue mentioned 
in Sec.~\ref{s:correlationC}). 

Next comes the difficult part of the analysis: 
one must divide the number of pairs per event 
by the number of uncorrelated pairs, as explained 
in Sec.~\ref{s:correlationB}. 
For a fixed orientation of the reaction plane $\Phi_R$, this 
number depends on $\Phi_R$ (see Sec.~\ref{s:correlationC}):
\begin{eqnarray}
\label{nuncorexact}
N_{\rm uncor}(\Phi_R)&=&2\pi \mean{N_1}p_1(\phi_1-\Phi_R)\cr
&&\times 2\pi \mean{N_2}p_2(\phi_2-\Phi_R). 
\end{eqnarray}
The $\Phi_R$-independent part, $\mean{N_1}\mean{N_2}$, can be 
obtained using a standard mixed-event analysis. 
The $\Phi_R$-dependent part, however, involves the (first) flow coefficients 
$v_n$ of both particles in the pair. 

To avoid this complication, the procedure suggested in 
Ref.~\cite{Heinz:2002au} is to used mixed events with 
{\it aligned event planes\/}. 
This procedure, however, is only approximate, because one 
mixes events with different {\it reaction\/} planes, although 
the estimated planes are the same. 
To be specific, let us compare in a simple case the distribution 
of uncorrelated pairs following the exact procedure, 
Eq.~(\ref{nuncorexact}), and using mixed events with 
aligned event planes. 
To simplify the calculation, we assume that only elliptic flow $v_2$ is 
present, and that it has the same value for both particles in the pair; 
we further assume that the pair angle is the mean angle, $x=\frac{1}{2}$ 
in Eq.~(\ref{defPhi}). 
Then the exact result is 
\begin{widetext}
\begin{equation}
\frac{N_{\rm uncor}(\Phi_R)}{\mean{N_1}\mean{N_2}}=
1+2 v_2^2\cos 2\Delta\phi
+4v_2\cos 2\Delta\phi\,\cos 2(\phi_{\rm pair}-\Phi_R)
+ 2v_2^2\cos 4(\phi_{\rm pair}-\Phi_R).
\end{equation}
This is to be compared with the result obtained following the method of 
Ref.~\cite{Heinz:2002au}:

\begin{equation}
\frac{N_{\rm mixed}(\Phi_R)}{\mean{N_1}\mean{N_2}}=
1+2 v_2^2\mean{\cos 2\Delta\Psi_R}^2\cos 2\Delta\phi
+4v_2\cos 2\Delta\phi\,\cos 2(\phi_{\rm pair}-\Phi_R)
+ 2v_2^2
\frac{\mean{\cos 2\Delta\Psi_R}^2}{\mean{\cos 4\Delta\Psi_R}}
\cos 4(\phi_{\rm pair}-\Phi_R),
\end{equation}
\end{widetext}
where $\Delta\Psi_R\equiv\Psi_R-\Phi_R$ 
is the difference between the estimated event plane and 
the true reaction plane. 
As expected, both results coincide when $\Delta\Psi_R=0$. 
Quite remarkably, the mixed-event method is correct to 
leading order in $v_2$ even when $\Delta\Psi_R\not=0$. 
However, it misses the  coefficients of order $v_2^2$. 
Typical values of the correction factors for the STAR 
experiment at RHIC are $\mean{\cos 2\Delta\Psi_R}^2\simeq 0.6$ and 
$\mean{\cos 2\Delta\Psi_R}^2/\mean{\cos 4\Delta\Psi_R}\simeq 1.3$. 

Besides the systematic uncertainty we just discussed, the price to pay 
for aligned mixed events is that one must essentially 
perform the whole correlation analysis for fixed values of both the pair 
angle $\phi_{\rm pair}$ {\it and\/} the estimated reaction plane $\Psi_R$. 
We suggest instead the following method:
\begin{enumerate}
\item place particle pairs in bins according to their rapidities $y_1$, $y_2$, 
total transverse momentum ${\bf K}\equiv {{\bf p}_T}_1+{{\bf p}_T}_2$, and 
relative momentum ${\bf q}$; 
\item in each such bin, build the correlation function $C({\bf q})$ as in 
the standard, azimuthally-insensitive HBT analysis; 
\item reconstruct the azimuthal distributions of pairs,  
$p(\phi_{\rm pair}-\Phi_R)$,
and of single particles, $p_1(\phi_1-\Phi_R)$ and  $p_2(\phi_2-\Phi_R)$, 
with respect to the {\em actual\/} reaction plane; 
that is, measure the first Fourier coefficients $v_n^{\rm pair}$ and $v_n$ 
for each particle in the pair (at RHIC energies, measuring the second and 
fourth harmonic, $v_2$ and $v_4$, should be enough to guarantee that the 
distributions are reasonably well reconstructed); 
\item with the help of the substitution Eq.~(\ref{subst}), build the 
azimuthal dependence of the correlation function. 
\end{enumerate}
One then eventually extracts azimuthally-dependent HBT radii 
using standard techniques (in particular, including correction for 
Coulomb effects) which are beyond the scope of this paper.

\subsection{Two-particle azimuthal correlations}
\label{s:jets}

Two-particle azimuthal correlations at large transverse momentum 
are under intense investigation in ultrarelativistic nucleus-nucleus 
collisions, since it has been realized that they yield 
direct evidence for hard scattering~\cite{Adler:2002ct}. 
In that case, one correlates a high-$p_T$ particle, the ``trigger'' 
particle, hereafter labeled 1, with a lower-$p_T$ particle, 
hereafter labeled 2. We assume for simplicity that particles 
1 and 2 belong to separate $p_T$ intervals. 
This is not the case in the STAR analysis~\cite{Adler:2002tq}, 
where particle 2 can be any particle with momentum lower than 
${p_T}_1$ above some cut. 
This difference, however, is not crucial for the following discussion. 

The following quantities must be measured: the average number of 
pairs per event, as a function of the relative angle $\Delta\phi$,
$\mean{N_{\rm pairs}(\Delta\phi)}$
(in practice, pairs are naturally sorted into equal-size bins 
of $\Delta\phi$), and the average numbers of particles per event 
$\mean{N_1}$ and $\mean{N_2}$. 

In studying the azimuthal dependence of the correlation, a natural 
choice is to take the azimuthal angle of the trigger particle $\phi_1$ 
as the pair angle $\phi_{\rm pair}$, i.e., one chooses $x=1$ in 
Eq.~(\ref{defPhi}). 
One needs to reconstruct the azimuthal distribution of pairs (for a given 
$\Delta\phi$ bin), $p^{\Delta\phi}(\phi_1-\Phi_R)$, and the azimuthal 
distributions of both trigger and associated particles, $p_1(\phi_1-\Phi_R)$ 
and $p_2(\phi_2-\Phi_R)$. 
One may then reconstruct the whole correlation function for a fixed value 
of $\phi_1-\Phi_R$. 
In particular, the correlation functions for the specific values 
$\phi_1=\Phi_R$ (in plane) and $\phi_1=\Phi_R+\pi/2$ (out of plane) are 
given by 
\begin{eqnarray}
C^{\rm out}(\Delta\phi)&\!\!=\!\!&
\frac{\mean{N_{\rm pairs}(\Delta\phi)}}{\mean{N_1}}\frac{p^{\Delta\phi}\!
\left(\frac{\pi}{2}\right)}
{p_1\!\left(\frac{\pi}{2}\right)} 
-2\pi c\mean{N_2}p_2\!\left(\frac{\pi}{2}+\Delta\phi\right)\cr
C^{\rm in}(\Delta\phi)&\!\!=\!\!&
\frac{\mean{N_{\rm pairs}(\Delta\phi)}}{\mean{N_1}}
\frac{p^{\Delta\phi}(0)}{p_1(0)}-2\pi c\mean{N_2}p_2(\Delta\phi),
\end{eqnarray}
where $c$ is a constant close to unity, as explained in 
Sec.~\ref{s:correlationA}. It is independent of $\Delta\phi$ 
and $\phi_1-\Phi_R$. 

Let us briefly compare the above procedure with the one suggested by 
Bielcikova et al.~\cite{Bielcikova:2003ku}: these authors show how to 
analyze correlations in- and out of an event plane, which is not 
exactly the reaction plane. Since the event-plane resolution is 
a detector-dependent quantity, this prevents quantitative comparisons 
between different experiments. In addition, the algebra to subtract 
the uncorrelated part is much simpler with our method.

\begin{figure}[tb!]
  \centerline{\includegraphics*[width=0.7\linewidth]{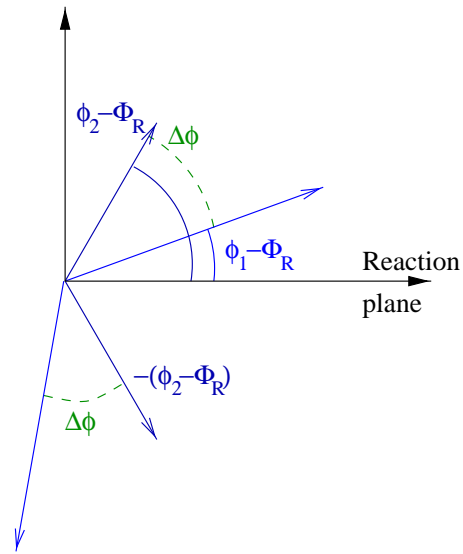}}
  \caption{Illustration of the prediction made in Eq.~(\ref{prediction}). 
    The long arrows represent the momenta of trigger particles, 
    while the shorter arrows represent the momenta of associated 
    particles. 
    If the modification of the correlation is due to the quenching of 
    the associated particle, it must be unchanged under 
    the transformation  $\phi_2-\Phi_R \to -(\phi_2-\Phi_R)$.
    \label{fig:jetquenching}}
\end{figure}

The standard interpretation of the modification of the correlation 
function in nucleus-nucleus collisions, compared to 
proton-proton collisions~\cite{Adler:2002tq}, is that 
the associated parton loses energy on its way through nuclear matter. 
If this interpretation is correct, then for a given $\Delta\phi$ 
the number of pairs per trigger particle depends only on the 
path followed by the associated particle. 
Symmetry with respect to the reaction plane implies that it 
is unchanged if $\phi_2-\Phi_R$ is changed into its opposite. 
As illustrated in Fig.~\ref{fig:jetquenching}, 
this symmetry is by no means trivial since the path followed 
by the trigger particle is now different.
This gives us, for arbitrary $\phi_2-\Phi_R$ and $\Delta\phi$, 
the prediction
\begin{equation}
\label{prediction}
\frac{p^{\Delta\phi}(\phi_2-\Phi_R-\Delta\phi)}
{p_1(\phi_2-\Phi_R-\Delta\phi)}=
\frac{p^{\Delta\phi}(-\phi_2+\Phi_R-\Delta\phi)}
{p_1(-\phi_2+\Phi_R-\Delta\phi)}. 
\end{equation}
If the only nonvanishing Fourier harmonic in 
the single-particle and pair azimuthal distributions 
is $v_2$, a simple calculation shows that the previous 
identity is equivalent to 
\begin{equation}
\label{predictionsinus}
v_{s,2}^{\rm pair}(\Delta\phi)=\left(v_{c,2}^{\rm pair}(\Delta\phi)-
v_2^{(1)}\right)\tan 2\Delta\phi, 
\end{equation}
where $v_2^{(1)}$ is the elliptic flow for the trigger particle, 
and the pair-flow coefficients 
$v_{c,2}^{\rm pair}$ and $v_{s,2}^{\rm pair}$ have been defined 
in Sec.~\ref{s:definitions}. 
This prediction is consistent with the general symmetry property
(\ref{property3}).

\section{Summary and perspectives}
\label{s:conclusions}

We have introduced novel, model-independent observables that describe the 
dependence in azimuth of two-particle correlations in heavy-ion collisions.
These observables, namely the coefficients $v_{c,n}^{\rm pair}$ and
$v_{s,n}^{\rm pair}$ in the Fourier expansion of the azimuthal distribution
(\ref{2p-dist2}) of the pair-angle $\phi_{\rm pair}$ that characterizes
(together with the relative azimuth) particle pairs, generalize in a natural
way the Fourier coefficients $v_n$ for single-particle anisotropic flow.
As the latter, the pair-flow coefficients can easily be measured in
experiments, using any ``usual'' method of flow analysis (modulo minor
modifications for the measurement of the sine terms, $v_{s,n}^{\rm pair}$): 
event-plane, two-particle correlations, cumulants, 
Lee-Yang zeroes can equally be applied. 
We however recommend the last two, when possible, in 
order to disentangle flow from non-flow effects. 

A main point of this paper is that these observables should replace, 
in future analyses, quantities that are defined for a given azimuth 
relative to the event plane. 
Much in the same way, the Fourier coefficients $v_n$ have now 
replaced earlier observables such as the ``flow angle'', the 
``squeeze-out ratio'' in most, if not all analyses of 
single-particle flow. 
The reason is that the event plane is not exactly the reaction
plane, and the dispersion varies from one experiment to the other. 
Therefore, observables defined with respect to an event plane only 
yield a qualitative information. 
The $v_n^{\rm pair}$, on the other hand, allow studies of 
azimuthally-sensitive correlations to enter the quantitative era. 

In a second part, we have briefly shown how to relate our observables 
to physical quantities of interest in three different cases: anisotropic
flow of decaying particles, interferometry, and azimuthal correlations of
high-momentum particles.
It is important to stress that, unlike the measurement of the pair-flow
coefficients $v_{c,n}^{\rm pair}$ and $v_{s,n}^{\rm pair}$, this second
step does depend on the underlying physical picture.
This model-dependence leads to some arbitrariness, which in practice
takes the form of the introduction of a normalization constant and (for
HBT and high-$p_T$-particle studies) a necessary truncation of the Fourier
expansion of the single-particle distribution.

A striking difference with usual studies of anisotropic flow 
is the general occurrence of sine terms in the Fourier series 
expansion. The relevance of such sine terms 
was already discussed in the context of single-particle flow 
\cite{Poskanzer:1998yz,Voloshin:2000xf} and azimuthally 
sensitive HBT~\cite{Heinz:2002au}. 
Here, we have shown for the first time that in the case of 
correlations between high-$p_T$ particles, jet quenching would 
result in a specific value for the sine term, given by  
Eq.~(\ref{predictionsinus}).

Azimuthally-sensitive correlations are among the subtlest analyses
in our field, and have already given valuable insight 
on the physics of high-energy nuclear collisions. 
We hope that the observables and methods introduced in this 
paper will help to improve future analyses. 
Thanks to the high statistics now available at RHIC, or 
that one can anticipate at the CERN Large Hadron Collider (LHC), 
new measurements will become possible, for instance the azimuthal 
dependence of non-identical-particle interferometry 
or the anisotropic flow of various ``new'' particle types, 
while probing new regions of phase space.

\begin{acknowledgments}
We thank Javier Castillo, Kirill Filimonov, Fran\c cois Gelis, 
Roy Lacey, Mike Lisa, Paul Sorensen and Raju Venugopalan 
for enlightening discussions, and Grzegorz Stefanek for useful 
comments on the manuscript. 
\end{acknowledgments}

\appendix
\section{Methods for analyzing single- and two-particle flow}
\label{s:appendix}

As stated in Sec.~\ref{s:definitions}, the measurement of the pair-flow
coefficients $v_{c,n}^{\rm pair}$ and $v_{s,n}^{\rm pair}$ involves the
same methods of analysis as for single-particle flow coefficients $v_n$ 
(modulo a small modification when measuring $v_{s,n}^{\rm pair}$).
This prompts us to recall briefly the various methods that have been proposed
in the literature, indicating the modification necessary to measure the 
sine coefficient. 

\subsection{Analysis of one-particle flow}
\label{s:methods1}

The most employed method of flow analysis at (ultra)relativistic energies
is the event-plane method~\cite{Danielewicz:hn,Ollitrault:1997di,%
   Poskanzer:1998yz}, which relies on the event-by-event determination
of an estimate of the reaction plane, the so-called ``event plane.''
Once the latter has been estimated (and various procedures to correct for
acceptance issues have been performed), one correlates its azimuth with
that of each outgoing particle, assuming that {\em all\/} correlations
between the event plane and a given particle (that is, actually, all
two-particle correlations) are due to flow. 
Eventually one must correct for the event-plane dispersion (computed with
the help of ``subevents'') on a statistical basis:
\begin{equation}
\label{vn-std}
v_n(p_T,y) = \frac{\mean{\cos n(\psi-\Psi_n)}}{\mean{\cos n(\Psi_R-\Phi_R)}},
\end{equation}
where $\psi$, $\Psi_R$, and $\Phi_R$ respectively denote the azimuths of the 
particle under study, the event plane, and the real reaction plane, while the 
denominator measures the event-plane dispersion. 

Beside this first method, it has long been known that anisotropic flow can
be analyzed with two-particle azimuthal correlations~\cite{Wang:1991qh},
without having to estimate the reaction plane in each event.
The procedure consists in building a two-particle correlator, similar to 
that employed in interferometry studies~\cite{Wiedemann:1999qn}, by forming 
the ratio of the number of ``real'' pairs (of particles in a same event) with 
relative angle $\Delta\phi$ over the number of ``background'' pairs (mixing 
particles from different events) with $\Delta\phi$:
\begin{equation}
C_2(\Delta\phi) \equiv 
\frac{N_{\rm pairs}(\Delta\phi)}{N_{\rm mixed}(\Delta\phi)}.
\end{equation}
As usual, dividing by ``mixed events'' automatically corrects for acceptance 
anisotropies, so that one can even work with a detector having only limited 
azimuthal coverage~\cite{Adcox:2002ms}, while the event-plane method requires 
an almost perfect azimuthal symmetry. 
The key point in constructing $C_2(\Delta\phi)$ is that its Fourier 
coefficients, which can be deduced by fitting the function, are precisely 
$\smean{\cos(n\Delta\phi)}=(v_n)^2$ and $\smean{\sin(n\Delta\phi)}=0$.
Letting first both particles in the pair run over the whole phase space 
covered by the detector, one obtains an estimate of ``integrated flow'', 
$v_n$, corresponding to some (detector-dependent) average of the coefficient. 
Restricting then one, and only one, of the particles in the pair (whose 
azimuth will be denoted by $\psi$) to some definite particle type, transverse 
momentum $p_T$ and rapidity $y$, while letting the other (azimuth $\phi$) be 
any particle in the event, one constructs a correlator whose Fourier 
coefficients are 
\begin{subequations}
\begin{eqnarray}
\mean{\cos n(\psi-\phi)} &\!=\!& v_n v_n(p_T,y) \label{vn-2cor}\\
\mean{\sin n(\psi-\phi)} &\!=\!& 0. \label{vn-2corbis}
\end{eqnarray}
\end{subequations}
The second identity reflects the evenness of $C_2(\psi-\phi)$ (when 
parity is conserved), while the first yields the ``differential flow'' 
$v_n(p_T,y)$. 
Please note that since only one particle per pair belongs to a small phase 
space bin while it is correlated to all other particles in the event, it 
follows that statistical errors are the same as with the event-plane method. 
Finally, the bias from ``nonflow'' effects~\cite{Ollitrault:dy} is of the 
same order of magnitude within both event-plane and two-particle correlation 
methods, but it is easier to subtract, when it is possible, in the latter, as 
exemplified in the case of unwanted correlations due to global momentum 
conservation in Ref.~\cite{Borghini:2002mv}. 

As already stated, the main limitation of both the event-plane and the 
two-particle methods is their relying on the assumption that all azimuthal 
correlations between particles result from their correlation with the 
reaction plane~\cite{Danielewicz:hn}.
In other words, they neglect nonflow correlations, whose magnitude is known
to be large at ultrarelativistic energies~\cite{Dinh:1999mn}.
One may try to subtract part of the nonflow effects by performing cuts in 
phase space, correlating together only particles that are widely separated 
(which certainly accounts for short-range correlations), but this results in 
larger statistical errors, while not removing entirely all unwanted effects.
The only systematic way to remedy the problem of nonflow correlations
in the flow analysis is to apply improved methods of analysis, based on 
multiparticle correlations~\cite{Borghini:2001vi,Borghini:2001zr,%
  Borghini:2002vp,Bhalerao:2003xf,Borghini:2004ke}, which have been 
implemented at the CERN Super Proton Synchrotron~\cite{Alt:2003ab} and at 
RHIC~\cite{Adler:2002pu}.
The essence of these methods is that the relative magnitude of nonflow
effects decreases, while that of collective anisotropic flow grows, when one 
considers the cumulants of correlations between an increasing number of
particles.
Measuring cumulants of four-, six-particle correlations, one thus minimizes 
the systematic error due to nonflow effects, the ultimate case being the use 
of Lee--Yang zeroes~\cite{Bhalerao:2003xf,Borghini:2004ke}, equivalent to 
``infinite-order cumulants,'' which isolate {\em collective\/} behaviors in 
the system, i.e., flow effects.
The price to pay is an increase in statistical uncertainties, but the latter 
is moderate at RHIC and LHC energies, especially if one uses all detected 
particles in the analysis.\footnote{One should not worry about possible double 
  particle counting, which amount to (unphysical) nonflow effects, and are 
  thus automatically taken care of in the methods.}

In practical analyses, these improved methods necessitate the computation of a 
generating function of multiparticle correlations, $G(z)$, where $z$ is a 
complex variable (see Eq.~(5) in Ref.~\cite{Borghini:2001vi} or Eqs.~(3),(5) 
in Ref.~\cite{Borghini:2004ke}). 
One then derives estimates of integrated flow: in the cumulant approach, by 
extracting the successive derivatives of $\ln G(z)$ at $z=0$ and identifying 
them with the corresponding derivatives of $\ln I_0(2v_n|z|)$~\cite{%
  Borghini:2001vi}; when using Lee--Yang zeroes, simply by looking for the 
location of the first zero of $G(z)$ in the complex plane (see Ref.~\cite{%
  Bhalerao:2003xf}, Eq.~(9)). 
Once estimates of integrated flow have been obtained, they are used to compute 
values of differential flow $v_n(p_T,y)$ for particles in a small $(p_T,y)$ 
bin, whose azimuthal angle we shall denote by $\psi$. 
This is done by correlating $\psi$ to the generating function (see Eq.~(26) in 
Ref.~\cite{Borghini:2001vi} or Eq.~(9) in Ref.~\cite{Borghini:2004ke}).

\subsection{Analysis of two-particle flow}
\label{s:methods2}

Any of the methods recalled in the previous Section can also be employed to 
measure the anisotropic flow of pairs as well, modulo small modifications. 
Whatever the method, the first step is strictly the same, namely the 
construction of the event plane (and the computation of its statistical 
dispersion) in the event-plane method, or the measurement of estimates of 
integrated flow in the two-particle and multiparticle methods. 

We shall now describe the changes that must be made to the measurement of 
differential single-particle flow in order to analyze the coefficients 
$v_{c,n}^{\rm pair}$, $v_{s,n}^{\rm pair}$. 
In short, the first necessary modification is the obvious replacement of 
$\psi$ (the azimuth of ``differential'' particles) by the pair angle 
$\phi_{\rm pair}$; then no further change is needed to obtain the cosine 
coefficient $v_{c,n}^{\rm pair}$, whose measurement strictly parallels that of 
$v_n(p_T,y)$, while for $v_{s,n}^{\rm pair}$ one should replace the 
``$\cos n\psi$'' term that is correlated to either the event plane or the 
other particles or a generating function with a ``$\sin n\phi^{\rm pair}$'' 
term.
Let us be more explicit:
\begin{itemize}
\item In the event-plane method, the pair-flow Fourier coefficients are given 
  by the averages 
  \begin{subequations}
  \begin{eqnarray}
    \displaystyle
    v_{c,n}^{\rm pair} &\!=\!& \frac{\mean{\cos n(\phi^{\rm pair}-\Psi_R)}}
    {\mean{\cos n(\Psi_R-\Phi_R)}}
    \label{vnc-std}\\
    v_{s,n}^{\rm pair} &\!=\!& \frac{\mean{\sin n(\phi^{\rm pair}-\Psi_n)}}
    {\mean{\cos n(\Psi_R-\Phi_R)}},
    \label{vns-std}
  \end{eqnarray}
  \end{subequations}
  where $\Psi_R$ is the event-plane azimuthal angle and the averages run 
  over pairs and events. 
  Note the analogy between Eqs.~(\ref{vn-std}) and (\ref{vnc-std}). 
\item When using two-particle correlations, one builds a two-point correlator 
  $C_2(\phi^{\rm pair}-\phi)$, where $\phi$ is any particle in the same event 
  as those involved in the pair, with the trivial exception of the pair 
  particles to avoid autocorrelations. 
  In opposition to the correlator used in single-particle flow studies, 
  $C_2(\phi^{\rm pair}-\phi)$ is no longer an even function, so that its 
  Fourier expansion has nonvanishing both cosine and sine coefficients which 
  are related to the pair-flow coefficients, namely
  \begin{subequations}
  \begin{eqnarray}
    \mean{\cos n(\phi^{\rm pair}-\phi)} &\!=\!& v_n v_{c,n}^{\rm pair} 
    \label{vnc-2cor}\\
    \mean{\sin n(\phi^{\rm pair}-\phi)} &\!=\!& v_n v_{s,n}^{\rm pair},
    \label{vns-2cor}
  \end{eqnarray}
  \end{subequations}
  where $v_n$ is the integrated flow while the averages run over all 
  $(\phi^{\rm pair},\phi)$ in each event, then over events. 
  Once again, Eq.~(\ref{vnc-2cor}) is reminiscent of Eq.~(\ref{vn-2cor}), 
  while the difference between Eqs.~(\ref{vn-2corbis}) and (\ref{vns-2cor}) is 
  due to the fact that whereas single-particle emission is symmetric with 
  respect to the reaction plane, pair emission, on the other hand, is not 
  symmetric, see Sec.~\ref{s:definitions} and Fig.~\ref{fig:angles}. 
\item To measure the Fourier   coefficient $v_{s,n}^{\rm pair}$ in the 
  Lee-Yang zeroes method, one should replace $\cos n(\psi-\theta)$ by 
  $\sin n(\phi_{\rm pair}-\theta)$ in the numerator of Eq.~(12) 
  (resp. Eq.~(9)) in Ref.~\cite{Bhalerao:2003xf} 
  (resp. Ref.~\cite{Borghini:2004ke}).%
\footnote{We assume for simplicity that $m$ takes the value 1 
in the cited equations. Values $m>1$ correspond to higher harmonics, 
for which the same modifications apply.}
\item Finally, in the cumulant method, the relevant cumulants when measuring 
  $v_{s,n}^{\rm pair}$ are the {\em imaginary\/} parts in the power-series 
  expansion of Eq.~(26-27) in Ref.~\cite{Borghini:2001vi}, while the real 
  parts are needed for the analysis of single-particle differential flow 
  $v_n(p_T,y)$ or of the cosine coefficients $v_{c,n}^{\rm pair}$. 
  As a result, the interpolation formula that allows one to extract the
  cumulants is similar to Eq.~(B7) of Ref.~\cite{Borghini:2001vi} (resp. 
  Eq.~(11) of Ref.~\cite{Borghini:2001zr}), modulo the replacement 
  $(X_{p,q}, Y_{p,q}) \to (Y_{p,q}, -X_{p,q})$, where $X_{p,q}$ and $Y_{p,q}$
  are still given by Eq.~(B6) in Ref.~\cite{Borghini:2001vi}.
\end{itemize}

Even if every method of single-particle flow analysis can in principle also be 
used to measure pair flow, modulo the modifications we described above, there 
exists a clear difference between the two-particle methods (both event-plane 
and two-particle correlation methods) on the one hand, and the multiparticle 
approaches on the other hand. 
As a matter of fact, we already mentioned that both two-particle methods rely 
on the assumption that all correlations between two arbitrary particles are 
due to anisotropic flow; in other words, that other sources of two-particle 
correlations are absent, or at most weak~\cite{Danielewicz:hn}. 
Now, if the purpose of using one of these methods is precisely to measure some 
azimuthally-dependent two-particle effect, the procedure is somehow 
self-contradictory. 
Such an inconsistency does not affect the measurement of pair flow through 
multiparticle methods, since the latter do not assume that two-particle 
correlations are inexistent, they merely minimize their effect.

\section{Symmetry properties of pair-flow coefficients}
\label{s:appendixB}

In this Appendix, we list a few mathematical properties of the pair-flow 
coefficients $v_n^{\rm pair}$ for the sake of completeness. 

The invariance of the two-particle distribution under the 
transformation $(\phi_1,\phi_2)\to(\phi_1+2\pi,\phi_2)$ 
translates into the (pseudo)periodicity property 
\begin{equation}
\label{periodicity}
v_n^{\rm pair}(\Delta\phi+2\pi)=v_n^{\rm pair}(\Delta\phi)\,\ee^{-2\ii\pi nx},
\end{equation}
where $x$ has been defined in Eq.~(\ref{defPhi}). 
If $x$ is changed to $x'$ in Eq.~(\ref{defPhi}), 
$v_n^{\rm pair}(\Delta\phi)$ is changed to 
$v_n^{\prime\,\rm pair}(\Delta\phi)
=v_n^{\rm pair}(\Delta\phi)\,\ee^{\ii n(x-x')\Delta\phi}$. 

If the system has symmetry with respect to the reaction plane (no parity 
violation), the two-particle distribution is unchanged under the joint 
transformation 
$(\phi^{\rm pair},\Delta\phi)\to (-\phi^{\rm pair},-\Delta\phi)$. 
At the level of the Fourier coefficients, this symmetry gives 
$v_n^{\rm pair}(-\Delta\phi)=v_{-n}^{\rm pair}(\Delta\phi)$. 
Together with property~(\ref{property1}), this yields
\begin{equation}
\label{property2}
v_n^{\rm pair}(-\Delta\phi)=v_{-n}^{\rm pair}(\Delta\phi)=
[v_n^{\rm pair}(\Delta\phi)]^*. 
\end{equation}
The corresponding properties for the 
real Fourier coefficients $v_{c,n}^{\rm pair}$ and 
$v_{s,n}^{\rm pair}$ are 
\begin{eqnarray}
\label{property3}
v_{c,n}^{\rm pair}(-\Delta\phi)&=&v_{c,-n}^{\rm pair}(\Delta\phi)=
v_{c,n}^{\rm pair}(\Delta\phi) \cr
v_{s,n}^{\rm pair}(-\Delta\phi)&=&v_{s,-n}^{\rm pair}(\Delta\phi)=
-v_{s,n}^{\rm pair}(\Delta\phi).
\end{eqnarray}
These various properties may prove useful to check that measured estimates 
of the Fourier coefficients behave ``properly''.

\end{document}